\documentclass[sigconf]{acmart}
\AtBeginDocument{%
  }

\acmConference[GI 2026]{Graphics Interface 2026}{June 09--12, 2026}{Waterloo, ON, Canada}
\acmBooktitle{Graphics Interface 2026 (GI 2026), June 09--12, 2026, Waterloo, ON, Canada}
\acmDOI{}
\acmISBN{}

\begin{document}

\title{AutoCue: Multimodal LLM-Assisted Externalization of Implicit Inputs as Instructional Visual Cues in Screencast Tutorials}

\author{Shengyang Luo}
\orcid{0009-0003-8707-7050}
\affiliation{%
  \institution{Purdue University}
  \city{West Lafayette}
  \state{Indiana}
  \country{United States}
}
\email{luo525@purdue.edu}

\author{Shengyao Luo}
\orcid{0009-0007-1481-7029}
\affiliation{%
  \institution{Virginia Tech}
  \city{Blacksburg}
  \state{Virginia}
  \country{United States}
}
\email{shengyao@vt.edu}

\author{Xiaolei Guo}
\orcid{0009-0002-8241-2078}
\affiliation{%
  \institution{University of Missouri}
  \city{Columbia}
  \state{Missouri}
  \country{United States}
}
\email{xiaolei.guo@missouri.edu}

\author{Fengze Zhang}
\orcid{0009-0002-6797-3862}
\affiliation{%
  \institution{Purdue University}
  \city{West Lafayette}
  \state{Indiana}
  \country{United States}
}
\email{zhan5455@purdue.edu}

\author{James Liang}
\orcid{0000-0002-6641-8077}
\affiliation{%
  \institution{Rochester Institute of Technology}
  \city{Rochester}
  \state{New York}
  \country{United States}
}
\email{jcl3689@rit.edu}

\author{Yingjie Victor Chen}
\orcid{0000-0001-6705-3535}
\affiliation{%
  \institution{Purdue University}
  \city{West Lafayette}
  \state{Indiana}
  \country{United States}
}
\email{victorchen@purdue.edu}






\renewcommand{\shortauthors}{Luo et al.}

\begin{abstract}
Tutorial videos are widely used for learning feature-rich software, yet following screencast tutorials often breaks down in practice. Through a survey and contextual inquiry, we found that learners frequently rewind or get stuck because critical input information, especially mouse actions and keyboard-modified operations, is often implicit or missing in tutorials without input metadata. To address this problem, we present AutoCue, a multimodal LLM-assisted, human-in-the-loop tutorial augmentation pipeline for externalizing implicit inputs as instructional visual cues. AutoCue integrates frame-to-frame visual changes, narration signals, and operation guidance from official software manuals to infer likely mouse and key-modifier actions, then produces aligned cue layers and editable artifacts for human refinement. Grounded in multimedia learning and cognitive load theory, we further develop a visual cue grammar for representing mouse, keyboard, and combined inputs in software-learning tutorials. We instantiate and evaluate AutoCue in Autodesk Maya, focusing automatic inference on selected UI-mediated interactions with observable visual or textual feedback while supporting more ambiguous state changes through editable authoring artifacts. In a between-subjects study with 24 participants, the AutoCue-augmented tutorial reduced task completion time and interaction breakdowns and showed improved learner-reported experience.
\end{abstract}



\begin{teaserfigure}
  \includegraphics[width=\textwidth]{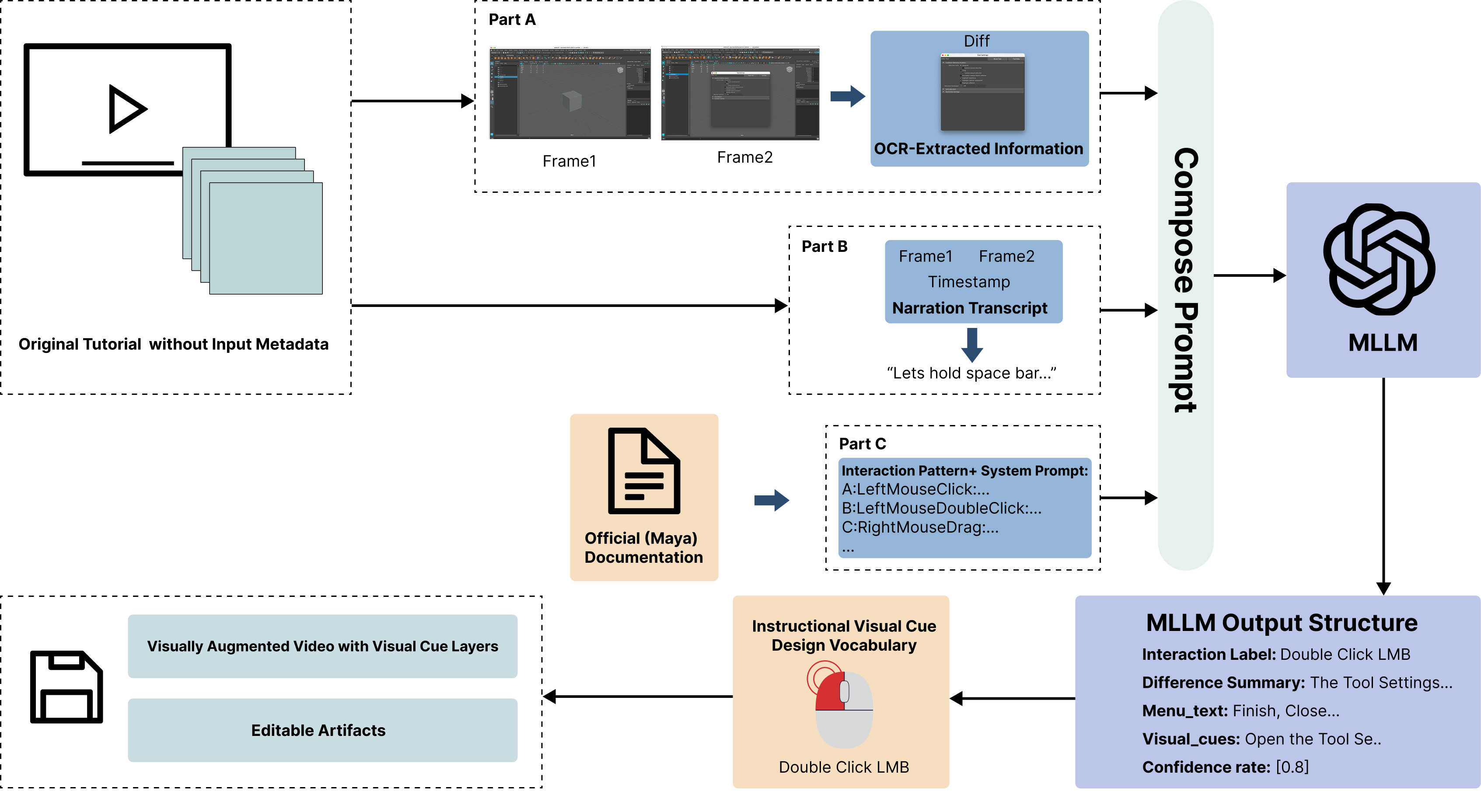}
  \caption{Overview of the AutoCue system workflow. AutoCue takes an original screencast tutorial without input metadata and analyzes three complementary evidence sources: frame-to-frame visual changes, narration transcripts, and interaction patterns derived from official software documentation. These inputs are composed into a constrained multimodal LLM prompt to infer implicit mouse, keyboard, and keyboard-modified actions. The inferred actions are then mapped to instructional visual cues and exported as both an augmented tutorial video and editable artifacts for human review and refinement.}
  \Description{}
  \label{fig:teaser}
\end{teaserfigure}


\maketitle

\section{Introduction}



Video-based tutorials are a main resource for learning feature-rich software, particularly visually intensive applications~\cite{yang2024aqua,kiani2019beyond}.
Among video-based tutorials, screencast tutorials provide step-by-step demonstrations of interface operations and action-reaction patterns that support software learning~\cite{kueker2024learning,van2016effects}.
Well-designed video tutorials can improve software learning outcomes, including procedural knowledge and self-efficacy~\cite{lloyd2012screencast,van2014comparison,mestre2012student}.
However, learners still frequently encounter frictions that manifest as repeatedly rewinding to rewatch segments for comprehension and becoming stuck during follow-along execution~\cite{yang2022softvideo,yang2024aqua}.
To improve follow-along learning and reduce friction, prior work has analyzed learners’ behaviors and developed tutorial-support systems that improve navigation, alignment, and in-video search~\cite{kim2014understanding,yang2022softvideo,chi2012mixt,fraser2020remap}. 
For example, Pongnumkul et al. developed Pause-and-Play, which synchronizes tutorial segments with application state~\cite{pongnumkul2011pause}. 
Banovic et al. presented Waken, which extracts interaction traces from software videos to support navigation~\cite{banovic2012waken}. 
Yang et al. introduced AQuA, which provides contextual question answering over tutorial videos~\cite{yang2024aqua}.
Despite these efforts in software learning, we still have a limited understanding of the usability barriers that underlie breakdowns and frictions in real-world software learning from screencast tutorials.


To better characterize these barriers and identify opportunities for support, we conducted two formative studies.
First, we surveyed learners’ practices and common frictions in authentic learning contexts and processes.
Consistent with prior research, learners commonly watch the tutorial while simultaneously operating the target software to follow along~\cite{kiani2019beyond,chi2012mixt}.
Participants reported common frictions, including frequent rewinding, becoming stuck, and obtaining different results despite following the same steps.
Building on these findings, we conducted a contextual inquiry to examine the usability barriers underlying these interaction breakdowns (rewinds, stuck events, and different results).
Across sessions, we observed that a major barrier was insufficient or implicit input information and visualization, such as mouse inputs ($e.g.,$ single click, double click, and click-and-drag) and keyboard-modifier input ($e.g.,$ hold Space + RightMouseButton drag), which learners had to infer on the fly during follow-along learning, increasing mental effort and contributing to friction, repeated rewinding, and getting stuck.

While some recent tutorials include metadata that display input actions ($e.g.,$  corner-based keystroke displays or click circles)~\cite{nguyen2015making,chi2014demowiz}, many widely used and popular tutorials, especially older ones, still lack clear input information because these cues typically require time-consuming manual authoring. 
Moreover, instructional cue designs vary widely across creators and platforms, leading to inconsistent visual encodings and information presentation. 
Without a coherent visual vocabulary and cueing grammar, learners may incur additional cognitive effort when switching between tutorials.

Motivated by these formative findings, we developed \textit{AutoCue}, a multimodal LLM-assisted, human-in-the-loop tutorial augmentation pipeline for software-learning screencast tutorials that lack input metadata. 
AutoCue uses multimodal LLM-based inference to recover implicit input information through an instructional visual cue vocabulary grounded in cognitive load theory and multimedia learning principles~\cite{mayer2005cognitive,mayer2002multimedia,ayres2005split}.
We instantiate and evaluate \textit{AutoCue} in Autodesk Maya, a visually intensive, feature-rich graphical user interface (GUI) application with frequent mouse, keyboard, and keyboard-modified interactions.
In this implementation, \textit{AutoCue} segments a tutorial into frames, analyzes frame-to-frame differences, and leverages an interaction menu derived from official Autodesk Maya documentation together with narration transcripts to infer likely input actions via a multimodal large language model, gpt-5.4 (see Figure~\ref{fig:teaser})~\cite{openai2026gpt54}.
The automatic output focuses on selected UI-mediated interactions with observable visual or textual feedback, such as menu selections, tool dialogs, and window transitions.
More ambiguous state or geometry changes are handled through editable artifacts, such as scripts or JSON files, allowing experts to add cues for critical interactions beyond UI-mediated events and refine the final augmentation in video production tools.

We evaluated the learner-facing value of \textit{AutoCue} in a between-subjects study with 24 Maya users by comparing the original tutorial against an AutoCue-workflow-augmented version.
Results showed that the workflow-augmented tutorial reduced task completion time and interaction breakdowns relative to the baseline. 
We conclude by discussing implications for instructional visual cue design, human-in-the-loop AI workflows, and future directions for adapting the architecture to other feature-rich software and tutorial contexts.


In summary, we make the following contributions:

\begin{itemize}
    \item \ An empirical formative investigation of frictions, interaction breakdowns, and usability barriers in follow-along learning from screencast tutorials, informing the design goals for the tutorial augmentation pipeline. 
    \item \textit{AutoCue}, a configurable, multimodal LLM-assisted, human-in-the-loop tutorial augmentation pipeline that automatically infers critical implicit inputs, particularly from UI-mediated interactions, and outputs both augmented videos and editable artifacts for expert completion and refinement.
    \item \ An instructional visual cue vocabulary and design grammar, grounded in multimedia learning principles, for externalizing mouse inputs (single clicks, double clicks, and click-and-drag actions) and keyboard-modified mouse inputs in screencast tutorials.
\end{itemize}

\section{Related Work}

\subsection{Video-Based Tutorial Design}
Prior studies show that video tutorials can outperform text-based materials in statistics learning and software procedure mastery, likely due to their multimodal presentation and learners’ ability to regulate pace and revisit difficult steps~\cite{lloyd2012screencast,van2014comparison,navarrete2025closer,saurabh2019modelling,brecht2008enabling}. As a result, video tutorials have increasingly displaced text-based resources across domains~\cite{lloyd2012screencast,van2014comparison,van2015test}. In this work, video-based tutorial learning refers to acquiring factual, conceptual, or procedural knowledge through audiovisual instructional videos~\cite{navarrete2025closer,fyfield2022improving,ten2015like}.

The growth of YouTube and other media-sharing platforms has expanded the creation and distribution of software tutorials, making screencast tutorials widely accessible~\cite{fyfield2022improving}. Screencast tutorials are especially common for software learning because they present step-by-step interface interactions and action sequences~\cite{van2016effects}. When designed according to multimedia learning principles, they can improve task performance, strengthen self-efficacy, and support retention over time~\cite{van2018supporting}.

Recent work has increasingly treated video tutorials as active learning tools rather than passive media. Researchers have therefore explored interactive elements such as clickable controls, contextual hints, and segment-based navigation to support engagement during viewing~\cite{van2018supporting,van2021practice}. Segmentation helps break complex procedures into manageable units and reduce cognitive load~\cite{garrett2021segmentation}, while reviewability supports targeted replay to reinforce understanding and retention~\cite{brar2017complex,van2016based,van2017reviews}. In \textit{AutoCue}, we reduce learners' action-inference burden by externalizing implicit input information as instructional visual cues to better support follow-along learning.

\subsection{AI and LLM in Video Tutorials}
Finding suitable tutorials remains challenging because learners often cannot express their immediate goals using the terminology assumed by tutorials and software features~\cite{drosos2024my}. Prior systems have addressed this challenge by progressively improving task-to-tutorial alignment. At the query level, QF-Graph bridges users’ task-oriented vocabulary and software functionality to support more effective tutorial search~\cite{fourney2011query}. RePlay brings search into the work context by using accessibility application programming interface (API) and captions to retrieve relevant clips across applications, reducing interruptions from switching to external search engines~\cite{fraser2019replay}. ReMap further lowers input burden during in-situ search by supporting multimodal queries through deictic pointing and speech~\cite{fraser2020remap}. More recently, Docent uses LLM reasoning to transform vague inputs and recent digital activities into novice-friendly, in-situ tutorials using operation-centric context~\cite{zhu2023docent}. WatchWithMe likewise uses LLMs to support interactive guided watching by generating transcript-based summaries, highlights, and question prompts during playback~\cite{chi2025watchwithme}.

Prior work has also expanded video-based tutorial learning from step alignment and navigation support to more recent forms of intelligent assistance. Pause-and-Play~\cite{pongnumkul2011pause} synchronizes screencast segments with application state so learners can access relevant steps without repeatedly shifting attention. For navigation, LectureScape~\cite{kim2014data} surfaces interaction peaks and ranks transcript hits based on aggregated viewing behavior, while Waken~\cite{banovic2012waken} visualizes user interactions on the timeline to support rapid seeking. Truong et al.~\cite{truong2021automatic} further showed that instructional videos can be transformed into hierarchical tutorials through computer vision and transcript analysis, enabling more structured navigation. Beyond navigation, MixT~\cite{chi2012mixt} and Community-Enhanced Tutorials~\cite{lafreniere2013community} explore content-level scaffolding by combining multiple modalities and community knowledge. More recent systems extend this direction with AI support. TutoAI presents an AI-assisted framework for mixed-media tutorial creation~\cite{chen2024tutoai}, while NoteIt converts instructional videos into interactive notes through multimodal video understanding~\cite{zhao2025noteit}. AQuA~\cite{yang2024aqua} uses an LLM-based interface for contextual question answering, and VideoMix applies a vision-language pipeline to aggregate information across multiple how-to videos to support broader task understanding~\cite{yang2025videomix}. In this paper, \textit{AutoCue} similarly leverages multimodal LLM reasoning to infer implicit input information and visualize it as instructional visual cues, supporting follow-along learning in software tutorials.

\subsection{Instructional Visual Cues}
Instructional visual cues, such as color highlighting, on-screen text, and arrows, are added visual information that scaffolds cognitive processing and directs attention within learning materials~\cite{de2007attention,mautone2001signaling}. Prior work on text-based and static illustrated materials shows that such cues can improve content recall and guide attention~\cite{cashen1970role,fowler1974effectiveness,hartley1985research,lorch1996effects}. Their design is commonly grounded in Mayer’s cognitive theory of multimedia learning~\cite{mayer1997multimedia,mayer2002multimedia} and Sweller’s cognitive load theory~\cite{sweller1988cognitive}. When cues guide attention spatially and temporally and are paired with prompts that elicit cognitive processing, they can improve understanding of animations and other dynamic materials~\cite{de2009towards}. At the same time, effective cueing can support comprehension of complex information while avoiding overload from excessive visual input~\cite{amadieu2011attention,semeraro2022visualizing}.

In instructional videos, visual cues are often categorized as textual, visual, or combined~\cite{wang2020impacts}. Beyond simply highlighting content, eye-tracking evidence suggests that such cues can shape how learners organize and integrate information during video-based tutorial learning~\cite{wang2020impacts}. They have therefore been adopted across diverse tutorial domains to support learning and improve user experience~\cite{semeraro2022visualizing,tang2015physio,ragazou2023effects}. For example, Semeraro and Turmo Vidal augment strength-training tutorials with post-produced cues such as arrows, line overlays, body highlights, and metaphorical 3D objects to make demonstrations clearer and more imitable~\cite{semeraro2022visualizing}. Ragazou and Karasavvidis similarly augment Blender tutorials with animated arrows and shapes that indicate key menus and options, helping learners follow procedural steps more efficiently~\cite{ragazou2023effects}. Aligned with cognitive load theory and multimedia learning principles, \textit{AutoCue}'s instructional visual cue design supports a structured visualization of interaction information.

\section{Formative Study 1: Survey}

To better understand how to support learning from screencast tutorials for feature-rich software, we conducted two studies on learners’ practices, device setups, and usability barriers. 
This paper reports only the findings that directly informed our system design. 
We first surveyed common learning practices and frictions, then conducted a contextual inquiry to examine the usability barriers underlying interaction breakdowns. 
Finally, we synthesized these findings into design space and goals that guided the development of our system to improve learning efficiency and experience.
This study, including the online survey, contextual inquiry, and evaluation, was approved by the Institutional Review Board (IRB) of Purdue University.
All procedures were conducted in accordance with institutional guidelines for research involving human participants.

\subsection{Survey Design}
To develop the survey, we first conducted a pilot study using semi-structured interviews (see Supplement 1.1) with five learners who had at least three months of experience learning feature-rich software through screencast tutorials. 
This criterion allowed us to recruit participants who had learning experience but were still actively learning through tutorials, helping us focus on practices and frictions in follow-along learning rather than difficulties driven mainly by general unfamiliarity. 
We then organized the interview insights through affinity mapping, which informed the design of the final survey questions (see Supplement 1.2).
 
\subsection{Participants}
We collected survey responses from 97 participants who had at least three months of experience learning feature-rich software through screencast tutorials. 
Most participants were undergraduate students in industrial design, engineering, and game development. 

\subsection{Results}
After excluding incomplete or low-quality submissions, we retained 87 valid responses for analysis. 
The average completion time was 3.74 minutes (SD = 2.26, Mdn = 3.57).

\subsubsection{Learning Method}
Drawing on the pilot study, we provided two main learning methods and an open-ended option. Among the 87 participants, most adopted a parallel learning strategy. 
Specifically, 87\% (n = 76) reported \textbf{following the tutorial while performing the steps in the software.} 
Only 10\% (n = 9) preferred \textbf{a sequential approach, watching the tutorial first and practicing afterward. }
The remaining 2\% (n = 2) reported other strategies, such as skipping to relevant segments and applying the demonstrated features directly to their own projects. 
Overall, these results indicate that most learners prefer to follow screencast tutorials step by step while working in the software.

\subsubsection{Common Frictions}
To capture common frictions during learning, we provided four options derived from the pilot study along with an open-ended "other" option. 
67\% (n = 58) reported \textbf{repeatedly playing back to specific moments to check the content and steps. }
A second major challenge was outcome mismatch, with 59\% (n = 51) reporting that they \textbf{followed the same steps as the tutorial but still obtained different results. }
Sustained engagement was also difficult, as 46\% (n = 40) reported \textbf{trouble maintaining attention during long tutorials. }
In addition, 39\% (n = 34) said that \textbf{switching between the video and the software disrupted their workflow by fragmenting attention. }
Only 6\% (n = 5) reported other issues, including version or hardware differences, difficulty dragging a touchscreen progress bar, and an overly fast pace.

\section{Formative Study 2: Contextual Inquiry}
Contextual inquiry is an in-situ qualitative method that combines observation and interviewing while users perform real tasks~\cite{raven1996using}. 
It is widely used in HCI to uncover hidden pain points and underlying usability barriers~\cite{krome2016contextual,li2024contextual,dosono2015m,de2004introducing,lischke2018understanding,jin2024virtual}.

To investigate the usability barriers underlying the survey findings, including frequent rewinding, outcome mismatch (getting stuck), and attention switching between the tutorial and the software, we conducted contextual inquiries in authentic learning settings. 
Because most survey respondents reported using video tutorials for 3D modeling, animation, or game-engine tools, we selected Autodesk Maya as a representative feature-rich 3D application.
Participants followed a step-by-step tutorial to model a 3D object while thinking aloud about their actions, reasoning, and difficulties. 
The participants in the survey and contextual inquiry were distinct.



\subsection{Participants}\label{clparticipants}
We conducted contextual inquiries with eight participants (7 male, 1 female) who had at least three months of Autodesk Maya experience. 
Participants’ mean age was 19.5 years (SD = 1.41) with an average of 4.9 months of Maya experience. 
All provided informed consent and received a small thank-you gift. 
Each contextual inquiry session lasted approximately 35 minutes.

\subsection{Tutorial Selection and Setting}
In the contextual inquiry, the tutorial served as a methodological probe to elicit usability barriers associated with the survey-reported challenges while approximating learners’ course- and project-based workflows. 
We selected an end-to-end object-modeling tutorial based on a set of criteria informed by prior instructional tutorial guidelines~\cite{van2022eleven} (see Supplement 2.1).
From five candidates sourced across YouTube, massive open online course (MOOC) platforms, and university course materials, we applied the rubric and chose the highest-rated option: a 10:35-minute Maya tutorial on modeling a low-poly 3D model, offering a realistic yet manageable workflow for examining usability barriers in learning from screencast tutorials.

\begin{figure}
    \centering
    \includegraphics[width=0.8\linewidth]{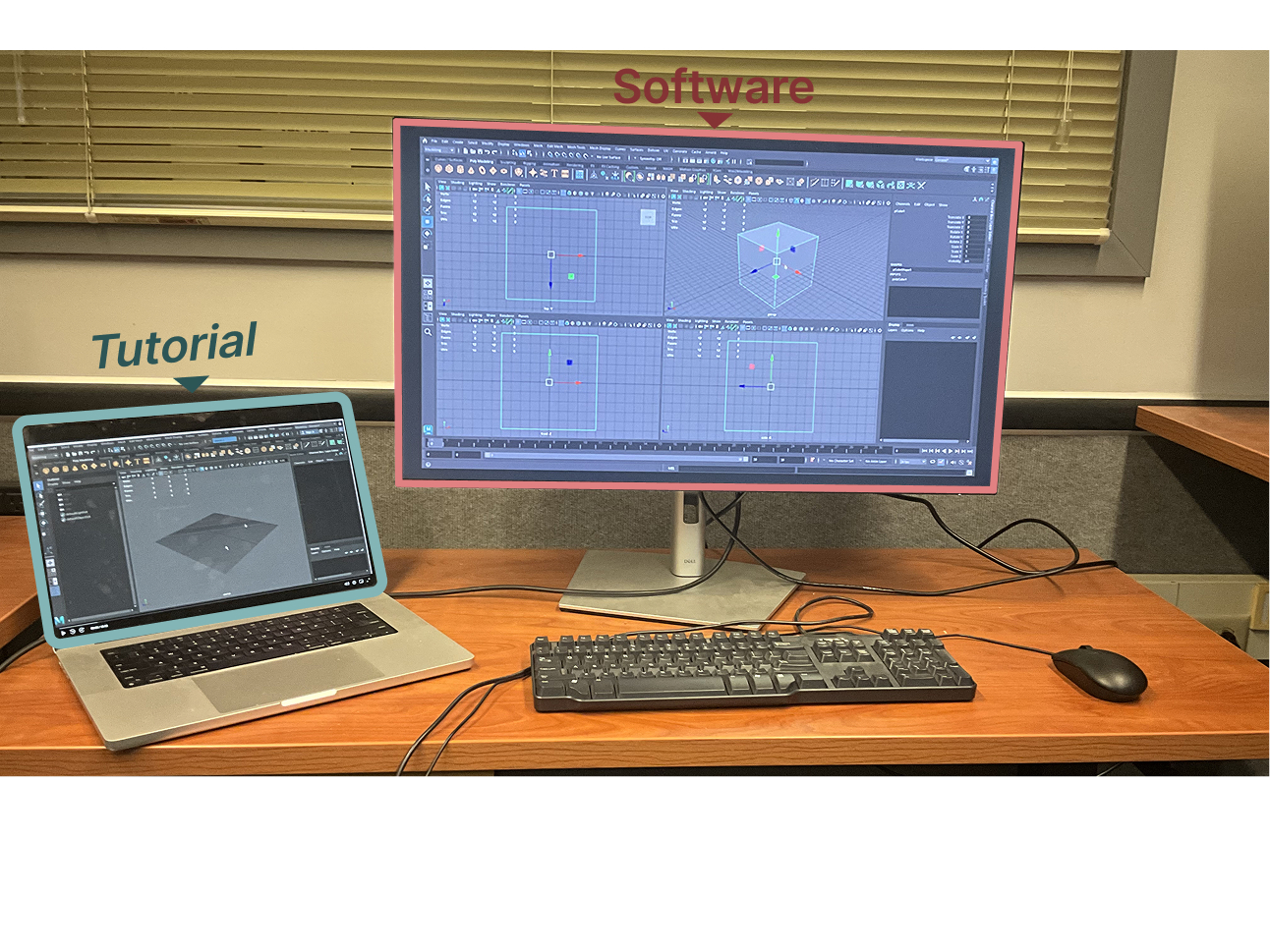}
    \caption{Contextual inquiry setup. Participants watched the screencast tutorial on a laptop while reproducing the demonstrated modeling task in Autodesk Maya on a separate monitor. This two-device setup allowed us to observe tutorial-control behaviors, such as pausing and rewinding, alongside learners’ software interactions with the keyboard and mouse.}
    \label{fig:annotator-a}
\end{figure}

We used a two-device setup: a Windows PC for the Maya task and a MacBook for playing the tutorial (Figure~\ref{fig:annotator-a}). 
This configuration made it easier to observe tutorial controls such as pausing, rewinding, and getting stuck. 
It also reflects learners’ everyday practice, as our survey identified it as the most common setup.
An instructor was behind the participant throughout the session. 
A tripod-mounted, high-resolution camera (smartphone-level quality) recorded the tutorial screen, the Maya interface, and mouse/keyboard hand interactions.

\subsection{Study Procedure}
The study comprised three components: a pre-study survey (5 minutes), a contextual inquiry (25 minutes), and a semi-structured interview (5 minutes).

In the pre-study survey, we collected participants’ demographic information (Section~\ref{clparticipants}).

During the contextual inquiry, participants followed the tutorial and independently reproduced the mug demonstration. 
Throughout the session, participants used a think-aloud protocol~\cite{van1994think} to verbalize their reasoning. 
When needed, the instructor asked brief follow-up questions to clarify actions.
The instructor also provided help when participants got stuck or could not move to the next step.

After the contextual inquiry session, we conducted a follow-up interview to clarify participants’ observed behaviors and to elicit their needs for screencast tutorials.
Interviews were audio-recorded.

\subsection{Analysis}\label{clanalysis}
To probe the usability barriers underlying our survey results, we analyzed pausing, rewatching, and getting stuck as breakdown patterns that consistently reflect frictions, including attention switching, repeated rewinding, and outcome mismatches.
We first used the Video Annotator tool~\cite{dutta2016via} to label each interaction breakdown along the tutorial timeline.
For each labeled event, we triangulated participants’ think-aloud comments with instructor observations and probing questions to infer the underlying cause and assign a reason code. 
Two researchers then independently conducted thematic analysis on these reason codes, iteratively clustering them into themes that represent usability barriers in tutorial design and screencast tutorial learning.

\subsection{Findings}
Our analysis suggests that pauses, rewinds, and stuck moments reflect different levels of usability issues. 
Pausing was usually low impact, as participants often resumed and completed later steps smoothly, consistent with prior work~\cite{mayer2002multimedia,merkt2018pauses}. 
Rewinding signaled greater friction that disrupted learning flow. 
Getting stuck was the most severe case, as participants could not proceed without help. 
Based on our thematic analysis, we summarize the reasons behind rewind and stuck events as distinct usability barriers in the learning process.

\subsubsection{Rewind During Learning}
\begin{figure}
    \centering
    \includegraphics[width=\linewidth]{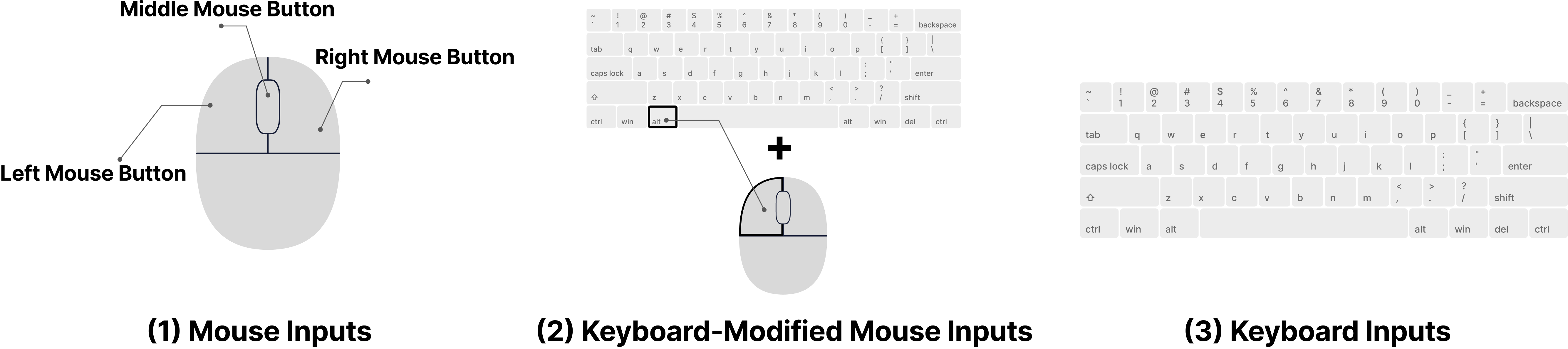}
    \caption{Three categories of implicit input information identified in the contextual inquiry. (1) Mouse inputs include single clicks, double-clicks, and drag actions; (2) keyboard-modified mouse inputs combine key presses with mouse actions; and (3) keyboard inputs include shortcuts and single-key commands. These input types were frequently missing or unclear in screencast tutorials and motivated the design of AutoCue’s instructional visual cue vocabulary.}
    \label{fig:hreeTypeAction}
\end{figure}

Participants often rewound because the tutorials provided insufficient visual and interaction cues for mouse, keyboard, and keyboard-modified mouse inputs. 
Based on our analysis, all participants reported rewinding after missing input actions. 
We grouped these inputs into three categories (Figure~\ref{fig:hreeTypeAction}): (1) Mouse inputs, (2) Keyboard inputs, and (3) Keyboard-modified mouse inputs.

\textbf{Keyboard inputs} mainly referred to hotkeys that the tutorial used but did not display on-screen. 
As with many feature-rich applications, Maya relies heavily on shortcuts ($e.g.,$ Ctrl+C/Ctrl+V) and also supports single-key bindings. 
For instance, the tutorial switched views using the spacebar, but participants did not recognize this the first time they encountered it (P2, P3). 
P2 noted that the tutorial was otherwise easy to follow, and that rewinds were mainly triggered by implicit shortcuts and keybinds.

\textbf{Mouse inputs} were also poorly cued in the videos, which prompted frequent rewinds. 
Although the mouse affords distinct buttons (left/middle/right) and common actions (click, double-click, click-and-drag), participants often could not tell which input the instructor used. 
This ambiguity repeatedly triggered rewinding (P1–P7).
One recurring case involved opening and configuring the Revolve tool: the instructor double-clicked to bring up the menu, but several participants interpreted it as a single click and failed to open the correct window (P3–P7).
Drag gestures created similar confusion. 
In Maya, dragging is essential for operations such as curve manipulation (drag left mouse button) and mode changes (drag left mouse button), yet participants sometimes treated drag as mere cursor movement (P1, P2, P3, P5). 
As a result, they hovered or clicked when the tutorial required a drag, which led to outcome mismatches and getting stuck. 
As P1 noted, the lack of visible mouse-input cues made it difficult to tell whether an action required a single click, double-click, or drag, and with which button.

\textbf{Keyboard–mouse combinations} were harder to follow because learners had to track two inputs at once. 
We observed that these moments frequently triggered rewinds and confusion (P1, P3, P4, P6, P8). 
For instance, P1 and P6 could not replicate the instructor’s action of holding Shift while left-clicking, even though it was described in the narration. 
In another case, P4 tried to change the view by right-dragging but missed that the instructor was also holding the spacebar, so the action failed. 
P8 noted that they often rewound because keyboard-modified mouse inputs were not clearly shown on screen. 
They added that although some tutorials display these cues, separating keyboard and mouse indicators across the screen still increases effort and often leads to repeated rewinding.

\subsubsection{Getting stuck during learning}
Most stuck cases occurred when learners thought they had replicated the instructor’s steps but obtained different outcomes. 
The main reason is that learners missed or performed extra steps based on the tutorial's demonstration.

During the contextual inquiry, participants most often became stuck after missing a demonstrated step or adding an unintended action. 
In both cases, they believed they were following the tutorial, but their actual input differed, producing unexpected outcomes (P1, P5, P7, P8). 
A common trigger was an unnoticed mouse action ($e.g.,$ a single click or double-click), which participants assumed they had performed, leading to an incomplete sequence and breakdown. 
In other instances, participants introduced extra operations (P6, P8) that changed object state and caused errors. 
For example, P6 selected an additional curve before applying Revolve, producing a distorted mesh. 
Similarly, P8 intended to select a curve and a face to extrude the handle, but inadvertently selected extra faces, resulting in a Maya display error due to an invalid operation.

\subsection{Design Space and Goals} \label{subsec:DG}
Based on our two formative studies, we distilled three design goals for our system to support learning from a screencast tutorial for feature-rich software. 
First, the system should externalize critical input information that is often implicit.
Second, it should provide an instructional visual vocabulary that visualizes interaction information clearly.
Finally, it should enable tutorial augmentation that is scalable, extensible, and easily adjustable to different creators and learning contexts.

\begin{itemize}
  \item \textbf{DG1: Externalize implicit inputs.} Detect mouse inputs, keyboard modifiers, and combined actions ($e.g.,$ clicks, drags, and combined interactions) and externalize them as instructional cues to reduce follow-along breakdowns ($e.g.,$ rewinds and getting stuck). 
  \item \textbf{DG2: Instructional visual cue vocabulary and grammar.} Develop an instructional visual vocabulary and cue grammar grounded in cognitive load and multimedia learning principles to present visual cues clearly and efficiently while providing appropriate scaffolding.
  \item \textbf{DG3: Scalable and configurable augmentation.} Develop a system that links input detection to interaction visualization, supports reuse across tutorial types, and provides an adjustable workflow for screencast tutorial creators.
\end{itemize}

\section{Instructional Visual Cue Design}\label{section:visualdesign}
To design a visual system that supports the augmented pipeline, we first refined the interaction primitives commonly used in feature-rich software.
Next, we surveyed existing instructional visual cue designs for screencast tutorials in software learning from YouTube, MOOC platforms, and other tutorial video sources. 
Finally, informed by findings from our contextual inquiry and interviews, we designed an instructional visual cue system that conveys interaction operations clearly and efficiently. 

Based on our contextual inquiry analysis (Section~\ref{clanalysis}), we categorize interactions into three primary types: (1) mouse interactions, (2) keyboard interactions, and (3) keyboard-modified mouse interactions.

\subsection{Current Instructional Visual Cue Design}\label{visualcuecurrentexample}


\textbf{Mouse Inputs Visualization.} We observed three common ways of visualizing mouse inputs in existing screencast tutorials (see Supplement 2.2). First, tutorials use cursor-following circle animations to indicate single and double clicks at the moment of interaction. Second, they show text labels, such as "Single click" or "Double click", in a fixed screen region. Third, they use icon-based cues that depict the mouse and highlight the activated button.

\textbf{Keyboard Inputs Visualization.} Keyboard inputs are presented more consistently across tutorials. Most designs use either text overlays showing the pressed key(s) or icon-based cues representing the shortcut.

\textbf{Keyboard-modified Mouse Inputs Visualization.} For combined keyboard-mouse interactions, input information is often presented in a fragmented way. In many cases, keyboard and mouse cues appear in different screen corners. In others, both are grouped as text labels in a single corner, such as "Wheel Down" or "Middle Mouse Button", which can still be hard to follow because the cue remains spatially distant from the cursor and the interaction itself.

\subsection{Design Grammar and Vocabulary}
Grounded in the cognitive theory of multimedia learning (CTML), which views working memory as capacity-limited, our design aims to redirect learners’ limited resources away from visual distraction and inefficient information processing toward model building~\cite{mayer2005cognitive}. 
In particular, instructional visual cues that reduce visual search and unnecessary attentional shifts can lower extraneous processing and free resources for organization and integration~\cite{de2009towards}.

\begin{figure}
    \centering
    \includegraphics[width=\linewidth]{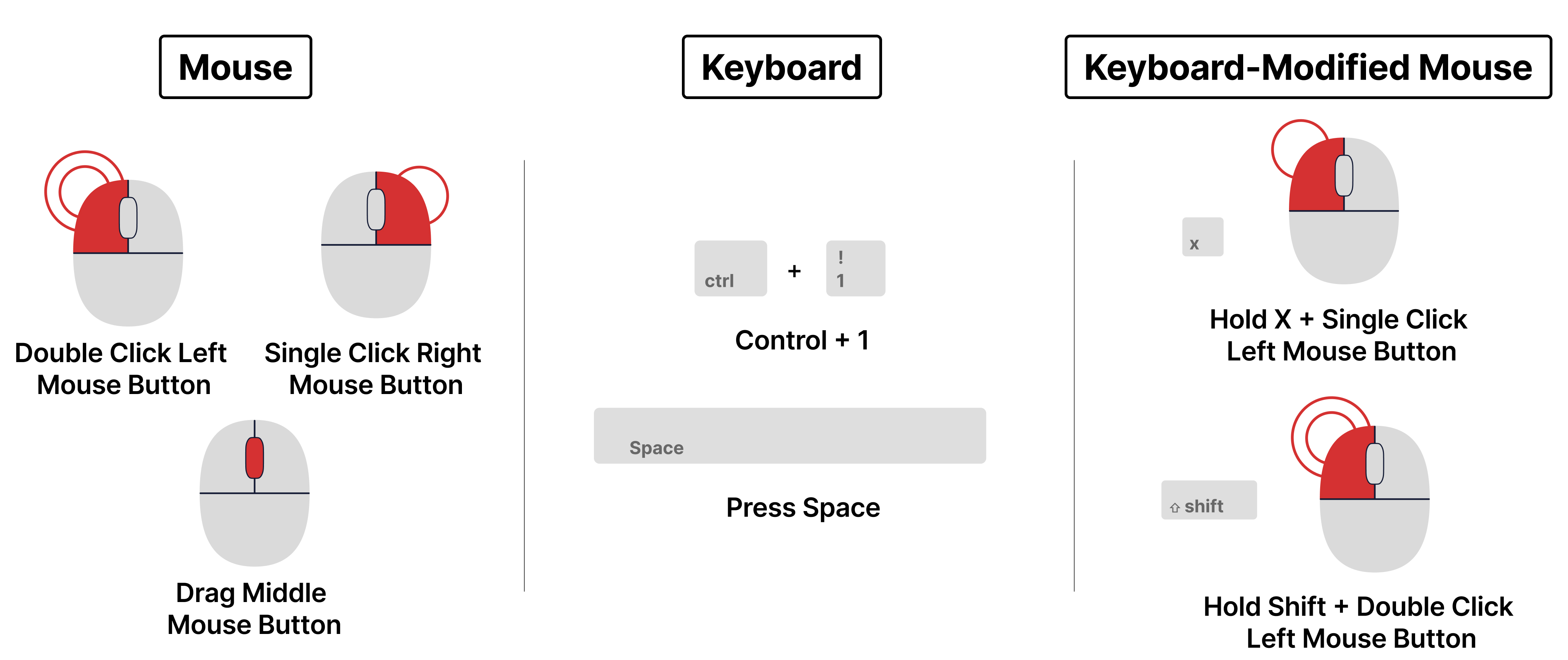}
    \caption{Instructional visual cue vocabulary for externalizing input actions. The vocabulary uses icon-based cues to represent mouse inputs, keyboard inputs, and keyboard-modified mouse inputs. Mouse-button highlights and click indicators distinguish actions such as left-click, right-click, double-click, and drag, while combined cues present keyboard modifiers and mouse actions as a single coordinated interaction.}
    \label{fig:DesignVocabulary}
\end{figure}

\textbf{Mouse Input Interaction.} To make mouse inputs explicit, we use icon-based graphics to represent mouse buttons (RMB, LMB, and MMB), and circular indicators to distinguish single clicks from double-clicks (Figure~\ref{fig:DesignVocabulary}).
This design externalizes input semantics at the point of interaction, reducing learners’ need to infer actions through visual search and lowering extraneous load~\cite{de2009towards,van2021signaling}.
Moreover, placing cues around the cursor minimizes gaze shifts between the cue and the cursor interaction position, consistent with the split-attention account~\cite{chandler1992split,ayres2005split}.
By improving spatial contiguity, the design further supports integration and schema construction~\cite{mayer2002multimedia}.

\textbf{Keyboard Input Interaction.} We display keyboard inputs in a clear bottom-screen overlay. This keeps key presses visible without obscuring the main workspace. 
Presenting key information in a consistent location also reduces extraneous processing associated with recalling or searching for implicit shortcuts~\cite{mayer2005cognitive}.

\begin{figure}
    \centering
    \includegraphics[width=\linewidth]{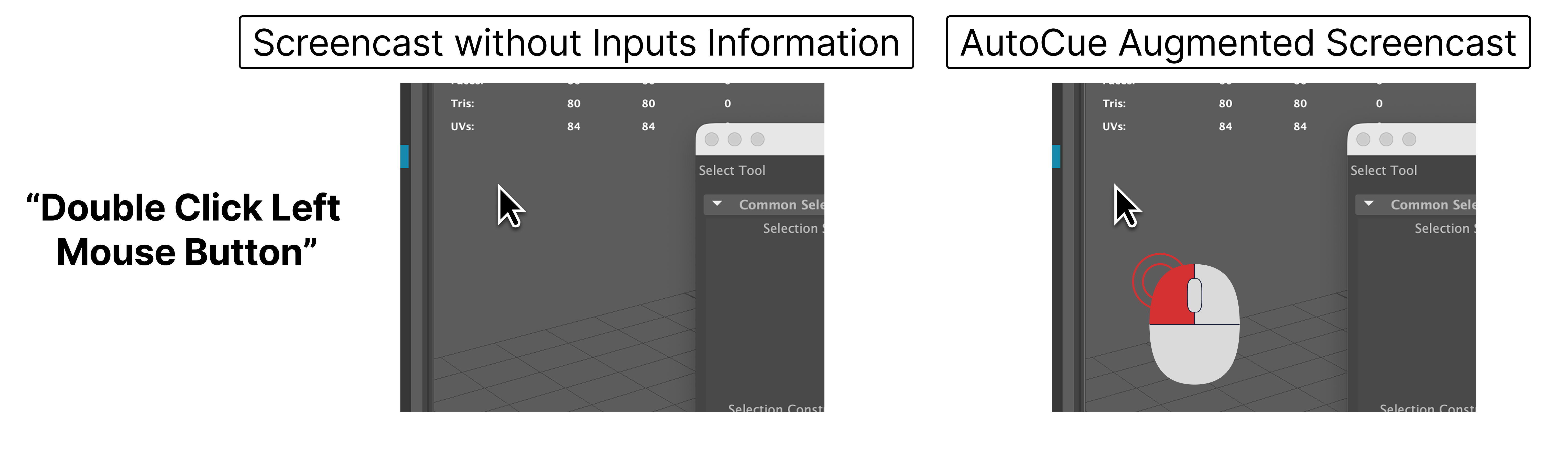}
    \caption{Before–after example of AutoCue augmentation. Left: the original screencast with implicit input. Right: the AutoCue-augmented screencast with an instructional visual cue for a left-mouse-button double-click. Autodesk screen shots reprinted courtesy of Autodesk, Inc.}
    \label{fig:Figure/ApplyinSoftUI}
\end{figure}

\textbf{Keyboard-modified Mouse Interaction.} For keyboard--mouse combinations, we present the key cue together with the mouse icon. 
This integrated representation helps learners perceive the coordinated input as a single composite action rather than linking two spatially separated cues through working memory~\cite{chandler1992split,ayres2005split}. 
Compared with common designs that separate mouse and keyboard cues (Section~\ref{visualcuecurrentexample}), our approach reduces cue distance, gaze shifts, and memory load. 
This design aligns with spatial contiguity principles and helps avoid split attention during follow-along learning~\cite{schroeder2018spatial,mayer2002multimedia}. Figure~\ref{fig:Figure/ApplyinSoftUI} compares the augmented tutorial with the original version.

\section{AutoCue}\label{sectionAutoCue}
Guided by our design goals, we present \textbf{AutoCue}, a human-in-the-loop tutorial augmentation system that infers and visualizes critical interactions from screencast videos.
\textit{AutoCue} addresses a key limitation of conventional software tutorials: many critical interaction details, such as mouse-button usage, click multiplicity, and modifier-assisted gestures, are visually subtle, transient, or entirely implicit.
As a result, learners often have to infer these actions through repeated rewinding, trial and error, or getting stuck.

\textit{AutoCue} implements a modular end-to-end pipeline that (1) identifies critical interaction moments from raw screencast footage and infers interaction inputs by combining visual change signals, on-screen text, and narration (DG1), (2) renders temporally and spatially aligned instructional visual cues that clearly convey input information (DG2), and (3) outputs either an augmented video or an editable authoring artifact for further refinement (DG3).
Rather than visualizing every low-level event ($e.g.,$ routine clicks or operations), \textit{AutoCue} prioritizes critical actions that are indirect or difficult to infer from the screencast tutorial and are frequently implicated in rewinding or getting stuck from our formative study. 
These include: (1) ambiguous click semantics (distinguishing single- vs.\ double-clicks and click-and-drag), (2) coordinated inputs (keyboard modifiers combined with mouse actions, such as hold Space + right mouse button drag), and (3) consequential events that produce substantive interface state changes.

\begin{figure}
    \centering
    \includegraphics[width=0.7\linewidth]{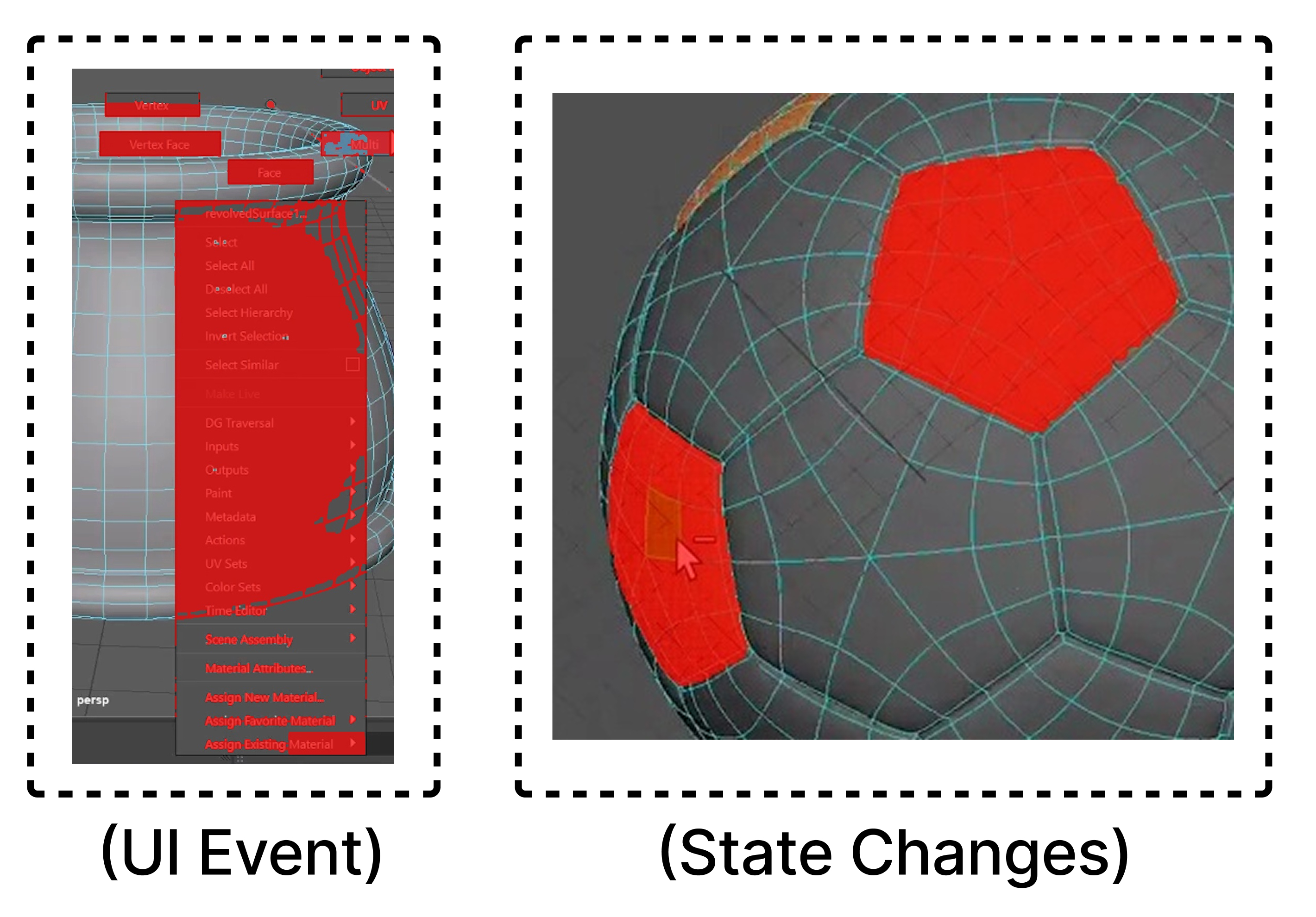}
    \caption{UI events vs. complex state changes. UI events provide explicit textual/iconic feedback (left), whereas complex actions manifest primarily as changes in geometry or selection state (right). Autodesk screenshots reprinted courtesy of Autodesk, Inc.}
    \label{fig:Figure/UIevent_ShapeChange_Vd}
\end{figure}

From an inference perspective, we broadly distinguish between two evidential manifestations of learner-relevant critical interactions in raw screencasts 
(Figure~\ref{fig:Figure/UIevent_ShapeChange_Vd}):
\begin{enumerate}
    \item \textbf{UI-mediated events:} actions with clear textual or iconic interface feedback, such as menu selections, tool dialogs, and window pop-ups.
    \item \textbf{Complex state changes:} actions expressed primarily through changes in geometry or object state, such as extruding, curve drawing, and multi-selection.
\end{enumerate}
To recover these interactions without input metadata or system logs, AutoCue uses constrained multimodal LLM-based inference over complementary evidence sources, including visual change, optical character recognition (OCR)-readable text, narration, and knowledge from official documentation.

Although our current implementation uses Autodesk Maya as a case study, AutoCue is designed to support adaptation to other feature-rich GUI applications. 
By separating interaction inference from cue rendering and parameterizing application-specific operation knowledge and label mappings, the pipeline could be extended to other software learned through screencast tutorials with observable visual feedback, such as Adobe Photoshop, Blender, and CAD tools.

\subsection{System Overview}
\begin{figure}
    \centering
    \includegraphics[width=\linewidth]{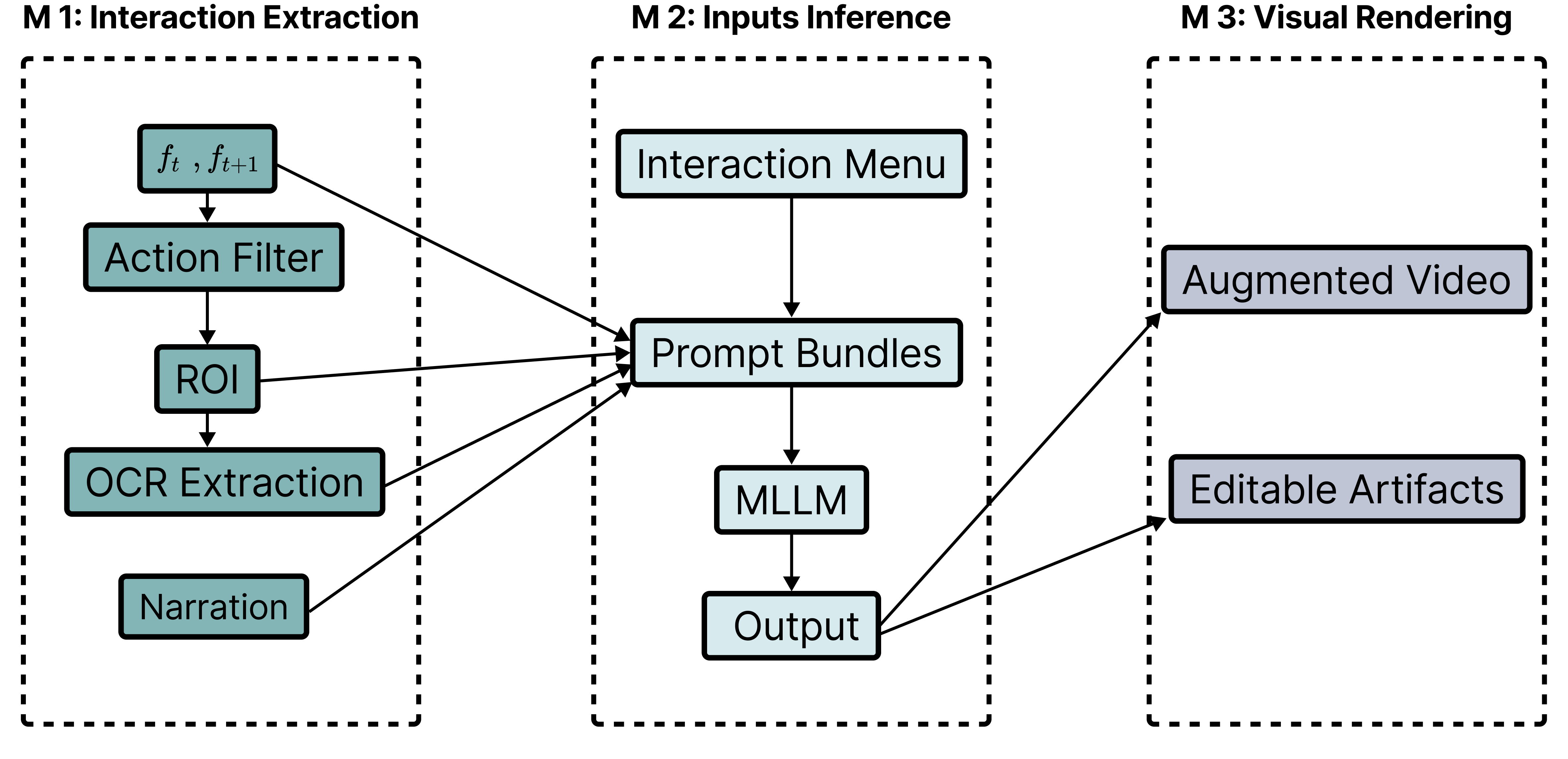}
    \caption{AutoCue architecture with three pipeline modules. M1 extracts candidate interaction moments from screencast footage by analyzing visual changes, OCR-readable interface text, and aligned narration. M2 performs constrained interaction inference by selecting input labels from a documentation-derived interaction menu. M3 renders the inferred inputs as instructional visual cues and exports both augmented videos and editable artifacts for expert refinement.}
    \label{fig:M1M2M3}
\end{figure}

As illustrated in Figure~\ref{fig:M1M2M3}, AutoCue is organized into three cooperating modules: \textbf{(M1) Critical Implicit Inputs Extraction}, \textbf{(M2) Constrained Interaction Inference}, and \textbf{(M3) Instructional Visual Cue Rendering and Export}. 
Given an input screencast $V=\{f_t\}_{t=1}^T$ with an optional narration track, the system first identifies a sparse set of candidate timestamps $\mathcal{K}\subset\{1,\dots, T\}$ that are likely to correspond to critical interaction moments. 
These timestamps serve as temporal anchors for subsequent inference.

For each candidate time $t \in \mathcal{K}$, AutoCue constructs a compact evidence bundle consisting of local frame context, regions of visual change, optical character recognition (OCR) output, and temporally aligned narration segments.
Interactions that produce explicit interface feedback, such as menus, tool dialogs, and window transitions, are particularly well suited to this representation.
These bundles are passed to a constrained multimodal LLM that selects interaction labels from a predefined interaction menu and returns structured descriptors with confidence estimates. 
Finally, the inferred interactions are translated into instructional visual cues that are spatially contiguous with the cursor and temporally synchronized with the original video, and exported either as an augmented tutorial video or as an editable project file for human verification and refinement.

The technical contribution of AutoCue lies not in novel low-level vision algorithms, but in integrating instrumentation-free interaction detection with multimodal evidence aggregation and constrained LLM-based inference, together with an authoring-aware output representation that supports human-in-the-loop refinement.

\subsection{Module M1: Critical Interaction Extraction} \label{subsec:M1}
Module~M1 identifies critical interaction moments and constructs evidence bundles for input information inference. 
The module operates directly on raw screencast tutorials and does not require access to application source code, input logs, or manual annotations.

\subsubsection{Difference-driven action filtering}

\begin{figure}
    \centering
    \includegraphics[width=\linewidth]{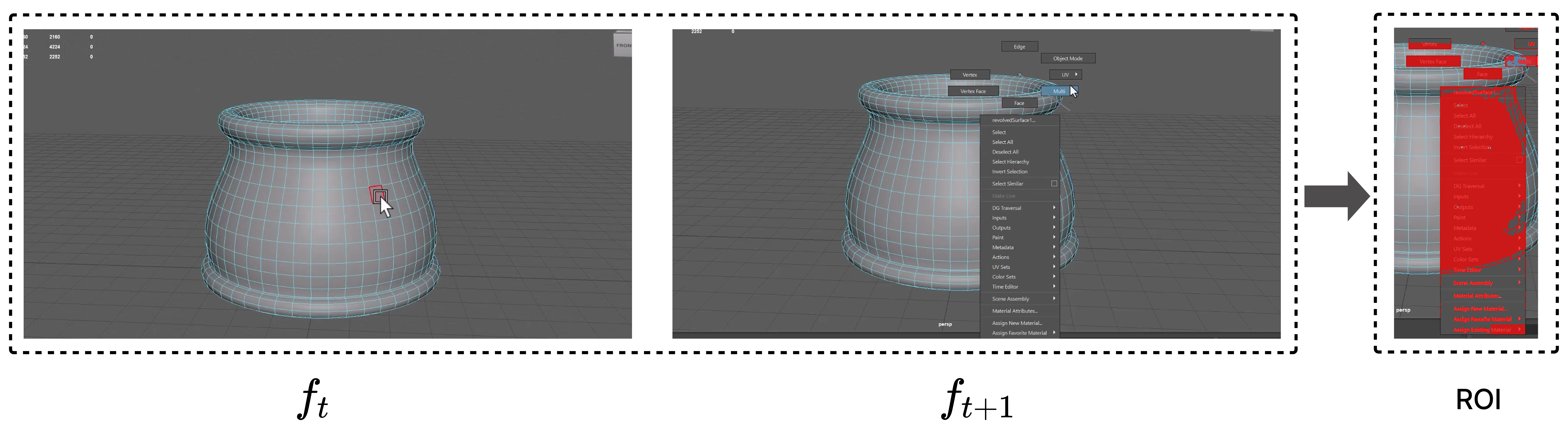}
    \caption{ROI extraction from adjacent frames. AutoCue compares $f_t$ and $f_{t+1}$ to localize a region of interest (ROI) containing salient UI changes (e.g., a newly opened menu). Autodesk screenshots reprinted courtesy of Autodesk, Inc.}
    \label{fig:Figure/F1F2ROI}
\end{figure}

AutoCue identifies candidate interaction moments via a difference-driven frame analysis. 
To infer inputs from visual feedback, we analyze the video frame sequence $\{f_t\}_{t=1}^{T}$ by computing a pixel-change score $d_t = \lVert f_{t+1} - f_t \rVert_1$\cite{radke2005image}.
This score drives an action filter that prioritizes key moments $t \in \mathcal{K}$ with significant structural variations ($e.g.,$ menu pop-ups, window transitions, shape changes), while suppressing low-signal segments like cursor jitter and general operations like movement/rotation to prevent overly dense outputs. 
For each selected moment, we extract the Region of Interest (ROI) containing the salient change and apply OCR (EasyOCR)~\cite{easyocr}, an off-the-shelf OCR engine independent of the multimodal LLM, to retrieve on-screen text context~(Figure~\ref{fig:Figure/F1F2ROI}). 
Finally, this textual evidence is combined with time-aligned narration scripts to serve as grounded references for the downstream LLM inference.

To identify candidate interaction moments while suppressing obvious noise, we combine pixel-wise Mean Absolute Difference (MAD) with Canny edge difference~\cite{canny2009computational}. 
Candidate processing is triggered when either signal exceeds its threshold, favoring recall at this stage. 
We then convert the resulting change responses into a binary mask using Otsu’s thresholding~\cite{otsu1975threshold} and clean it with morphological operations~\cite{serra1983image}. 
Finally, we apply simple geometric heuristics to deprioritize broad navigation and isolated cursor jitter, retaining moments that are more likely to correspond to meaningful interface or structural changes for downstream inference.


\subsubsection{OCR grounding}
To capture explicit interface feedback, AutoCue applies OCR to each ROI and to the surrounding frame region. 
The extracted tokens include menu content, window names, parameter values, and other transient UI text that often co-occurs with interaction events. 
Each token is stored together with its bounding box and timestamp, forming a grounded textual representation $\mathcal{O}_t$ that is later used as evidence during interaction inference.

\subsubsection{Narration alignment}
Instructors usually emphasize interaction details that are difficult to infer visually, such as double-click or keyboard-modified mouse inputs. 
AutoCue transcribes the narration track using an automatic speech recognition system and segments it into timestamped utterances. 
For each candidate interaction time $t$, the system aligns nearby narration segments within a fixed temporal window, producing a narration context $\mathcal{N}_t$. 
This multimodal alignment supports the disambiguation of visually similar interactions and improves robustness in visually complex scenes.

\subsection{Module M2: Implicit Input Inference} \label{M2}
Module~M2 infers interaction inputs from the evidence bundles produced by Module~M1. 
Rather than generating free-form textual descriptions, AutoCue constrains inference to a predefined interaction menu derived from official Autodesk Maya documentation. 
Constraining the output increases consistency, facilitates downstream rendering, and enables confidence-scored human review.

\subsubsection{Interaction menu}
AutoCue employs an interaction menu (see Supplement 3.1) derived from official Maya documentation and enumerates common input primitives and patterns, including mouse actions ($e.g.,$ single-click, double-click, drag), and modifier-assisted mouse interactions. 
During inference, the LLM is instructed to select exactly one interaction label from this menu for each candidate timestamp, or to return a "No interaction" if no meaningful interaction is present.
This constraint improves output consistency, reduces ambiguity, and ensures compatibility with downstream visualization.

\subsubsection{Constrained prompting and structure}
For each candidate $t \in \mathcal{K}$, the LLM receives a compact prompt containing (1) local frame pair ($f_{t}$, $f_{t+1}$) as context information with ROI, (2) OCR tokens $\mathcal{O}_t$, (3) narration context $\mathcal{N}_t$, and (4) the interaction menu. 
The model outputs a structured interaction descriptor that includes the selected label, ROI content description, a brief evidence-based rationale, and a confidence score in $[0,1]$ (See example in Supplement 3.2). 
These descriptors form a temporally ordered interaction sequence suitable for both rendering and human review.

\subsection{Module M3: Visual Rendering and Export}
Module M3 translates structured interaction descriptors into instructional visual cues and exports them in formats suitable for learning and authoring.

\subsubsection{Instructional Visual Cue Rendering}
Instructional cues are realized as a visual cue vocabulary with compositional rules (Section~\ref{section:visualdesign}). 
Mouse-related cues are rendered near the inferred cursor location to maximize spatial contiguity and minimize gaze shifts. 
Temporal persistence and visual prominence are configurable to balance informativeness and visual simplicity.


\subsubsection{Output formats}
AutoCue produces two complementary outputs:
\begin{enumerate}
  \item \textbf{Automatic augmented video}, instructional visual cues are overlaid directly onto the original screencast for learner consumption.
  \item \textbf{Editable authoring artifact}, such as a script or structured JSON file (see Supplement 3.3), where each cue is represented as an editable layer linked to its timestamp, position, evidence, and confidence score.
\end{enumerate}

To avoid introducing misleading cues in learner-facing augmentation, the automatic video includes only high-confidence critical interactions.
We define high-confidence interactions as predictions with a model-reported confidence score of at least 0.80, supported by salient ROI changes, OCR-readable interface text, and consistent with the documentation-derived interaction labels.
These cases are typically UI-mediated events, such as menu selections, tool dialogs, and window pop-ups, where the interaction produces explicit visual or textual feedback.
Interactions with weaker evidence and predictions below this threshold are retained in the editable artifact for review and refinement. 
In particular, complex state changes are often expressed through continuous changes in geometry or selection state, and therefore usually require human review to refine their interpretation and timing. 
By contrast, UI-mediated events provide explicit interface feedback at the moment of interaction, making them both instructionally important and more directly recoverable from AutoCue’s evidence-bundle formulation.

For both outputs, AutoCue also generates a CSV file that records each event’s timestamp, predicted label, supporting evidence, confidence score, and OCR-extracted text. 
This dual-output design supports scalable automation while preserving flexibility for correction and broader coverage through human refinement.

\subsubsection{Implementation}

In our implementation, we accessed the multimodal LLM through the OpenAI API using the Python SDK and used \texttt{gpt-5.4}~\cite{openai2026gpt54} with temperature set to 0 for more consistent outputs.
For each candidate timestamp, the model received two consecutive frames, a difference-highlighted ROI crop, and available OCR text and narration transcripts.
It returned a structured JSON prediction with the interaction label, confidence score, visual-change summary, and supporting evidence.
The full prompt and an example output are provided in Supplement 3.2.

\subsection{Technical Validation}\label{technicalvalidation}
We evaluated AutoCue on in-the-wild Maya tutorials to assess its robustness and practical utility as an augmentation pipeline. 
Because critical interactions vary in how directly they can be inferred from raw screencasts, this section focuses on UI-mediated events. 
These events provide explicit interface feedback, occur frequently in Maya tutorials, and account for a substantial portion of learner-relevant critical steps. 
They are, therefore, the interaction class best suited to automatic augmentation, while broader and more ambiguous cases remain supported through the editable workflow.

\subsubsection{Dataset and Procedure}


We created a dataset of 30 video segments (totaling approximately 45 minutes) derived from 5 distinct Maya tutorials for modeling by different creators, tasks, and UI layouts (see Supplement 4.1). 
The validation dataset was scoped to Maya modeling workflows, aligning with the formative and user-study tasks.
The sampled clips captured a range of modeling-relevant interaction patterns, including curve creation, revolve operations, mesh editing, view manipulation, tool-option dialogs, and selection-related UI feedback.
Together, these clips provide coverage of common interaction types in Maya modeling tutorials while preserving a focused evaluation scope.
Using a stratified random sampling strategy, we extracted six 1--2 minute clips from the beginning, middle, and end of each tutorial to capture variation in pacing and workflow complexity.

Two expert Maya instructors manually annotated learner-relevant critical interactions in the sampled clips. 
We then processed these clips through \textit{AutoCue} and compared the generated cues against the ground truth. 
Across the annotated clips, UI-mediated events approximately accounted for 40\% of learner-relevant critical steps. 
For technical validation, we focused on a modeling-specific subset of Maya interactions and operationalized six recurrent input and menu families, which expanded into more than ten concrete menu realizations in our annotated clips (see Supplement 4.2).
All quantitative performance metrics reported below were computed over this UI-mediated tier.

\subsubsection{Results: Coverage and Precision}

Across all validated events, AutoCue achieved a micro-averaged recall of 95.10\% and precision of 96.91\%.
The per-video standard deviation was 3.87\% for recall and 3.98\% for precision (Table~\ref{tab:recall_precision}). 
AutoCue performs reliably on UI-mediated events, which provide strong observable evidence in screencast tutorials. 
Because these events often induce salient pixel-level transitions and include OCR-readable interface text ($e.g.,$ menus and window titles), our difference-based detector and downstream LLM inference can localize and label them with high fidelity, substantially reducing manual annotation effort for routine UI interactions.

\begin{table}[t]
\centering
\caption{Recall and precision across five evaluation clips. Recall is computed as correct predictions over validated events, and precision as correct predictions over predicted events. The Avg. row reports micro-averaged results across clips.}
\label{tab:recall_precision}
\begin{tabular}{lcccc}
\toprule
ID & Recall (raw) & Recall & Precision (raw) & Precision \\
\midrule
V1 & 28/29 & 96.55\% & 27/28 & 96.43\% \\
V2 & 15/15 & 100.00\% & 15/15 & 100.00\% \\
V3 & 22/23 & 95.65\% & 20/22 & 90.91\% \\
V4 & 15/16  & 93.75\% & 15/15   & 100.00\% \\
V5 & 17/19 & 89.47\% & 17/17  & 100.00\% \\
\midrule
Avg. & -- & 95.10\% & -- & 96.91\% \\
\bottomrule
\end{tabular}
\end{table}

\subsubsection{Qualitative Error Analysis}
We conducted a qualitative analysis to understand the discrepancies between the system output and expert ground truth.

\textbf{Why recall is not 100\% (Missed Cues):}
Some missed cues result from AutoCue’s sampling-based pipeline. 
To keep multimodal inference tractable, we sampled frames at 0.5-second intervals, which could miss brief UI events when they occurred between sampled frames. 
This issue was more likely in tutorials containing sped-up segments or low-quality recordings, where menus became too brief or too blurry to support reliable OCR grounding and interaction inference.
Future improvements could reduce such misses through denser or adaptive frame sampling.

\textbf{Why Hallucinations Occur (Precision Gaps):}
Some precision errors arise when the same menu or dialog can be opened through different interaction paths. 
In these cases, AutoCue may observe a similar interface outcome but infer the wrong triggering action, because the current system does not yet explicitly model alternative ways of reaching the same UI state. 
Future improvements that incorporate stronger step-level context may help disambiguate these cases.

\section{User Study}
We evaluated the learner-facing value of the AutoCue workflow and instructional visual cue design using an Autodesk Maya screencast tutorial, as Maya is a visually intensive, feature-rich application. The tutorial demonstrates how to model a cup in Maya and was recorded without mouse or keyboard-modified mouse input metadata, representing a raw tutorial setting. It was newly created by an instructor with over 20 years of Maya teaching experience to reduce the likelihood that participants had previously seen it.

To construct the augmented tutorial, we processed the raw tutorial through AutoCue. Among 24 learner-relevant critical interaction steps identified in the tutorial workflow, AutoCue automatically generated cues for 13 UI-mediated events. Consistent with our human-in-the-loop authoring workflow, the expert instructor then reviewed these cues and completed the remaining 11 complex state changes to ensure that the final augmented tutorial was complete, pedagogically coherent, and did not mislead learners.

We compared the original tutorial against this AutoCue-workflow-augmented tutorial to examine two questions: (1) whether the instructional visual cue design effectively supports follow-along learning, and (2) whether the workflow can produce a learner-facing augmented tutorial that improves the learning experience. 
Using a between-subjects design, we measured task completion time, interaction breakdowns (rewinds and stuck events), and subjective learning experience and mental effort using a 7-point Likert-scale survey, since a within-subject comparison would be confounded either by prior exposure to the same workflow or by differences across alternative workflows.
We also conducted a semi-structured interview to collect qualitative feedback on the learning experience (see Supplement 5.1).

\subsection{Participants}


We recruited participants with at least three months of prior experience learning and using Autodesk Maya to be able to follow along the tutorial with fundamental knowledge and evaluated the modeling tutorial under two conditions (original tutorial and AutoCue-augmented tutorial). 
In total, 24 participants (10 female, 14 male; mean age = 21) were randomly assigned to one of two groups.
Participants included both undergraduate and graduate students majoring in industrial design, game development, and engineering.
No one attended the previous contextual inquiry or survey.

\subsection{Task and Measures}
Participants used a two-device learning setup that mirrors common real-world practice. 
A MacBook laptop played the tutorial video, and a Windows PC was used to follow the tutorial and perform the task in Maya.
An instructor sat behind each participant to record interaction breakdowns, specifically rewinds and stuck events, during the task.
The task included three stages (see Supplement 5.2): (1) drawing a curve with the Bezier tool; (2) revolving the curve into a mesh to form the mug body; and (3) refining the model by adjusting proportions and adding a handle.
When a participant became stuck and could not proceed, the instructor provided assistance to allow the session to continue.

We used the original tutorial as the baseline to isolate the effect of cue augmentation without introducing confounds from differences in tutorial quality, instructor style, narration, or pacing. 
Rather than serving as a weak comparison condition, the baseline was a usable raw screencast tutorial created by an experienced instructor, allowing us to test whether AutoCue could provide additional support beyond an already pedagogically coherent tutorial (see Supplement Video).

After completing the modeling task, participants then completed three 7-point Likert items assessing learning experience and mental effort. 
These items were informed by established subjective measures of mental effort and cognitive load in screencast tutorial or video-based tutorial learning research, such as NASA-TLX dimensions~\cite{hart1988development,klepsch2017development,ragazou2023effects,yang2024aqua}.
Finally, participants in the AutoCue condition provided qualitative feedback, including usefulness, suggestions, and design recommendations for the instructional visual cue design and the AutoCue system.


\subsection{Results}


We used two-sided Mann--Whitney U tests because the study used a between-subjects design with two independent groups and a small sample size ($N=12$ per condition).
Given this sample size, we did not assume normally distributed completion times or breakdown counts, and the Likert-scale ratings were treated as ordinal measures.
We therefore report medians and interquartile ranges as descriptive statistics, use Cliff's $\delta$ to describe effect magnitude, and report Hodges--Lehmann estimates for task completion time and breakdown counts to describe the estimated shift between conditions.

\subsubsection{Task Performance and Interaction Breakdowns}

\begin{figure}
    \centering
    \includegraphics[width=\linewidth]{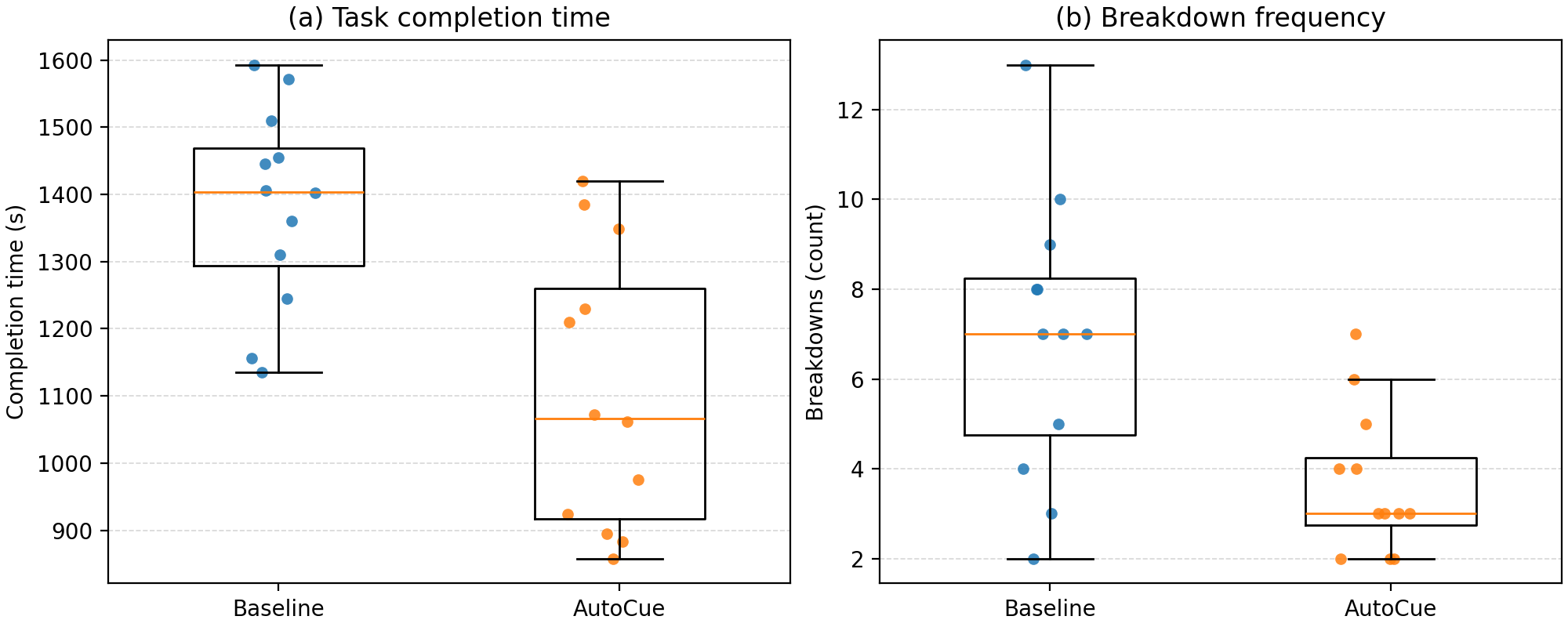}
    \caption{Task completion time and breakdown frequency for the Baseline and AutoCue conditions. Boxes indicate IQRs, center lines indicate medians, and points indicate individual participants (N=12 per condition).}
    \label{fig:timeandbreak}
\end{figure}

Task completion time differed significantly between conditions ($U = 20$, $p = 0.003$). 
Participants in the AutoCue condition completed the task substantially faster (median = 1067~s, interquartile range (IQR) = [916.8, 1259.8]) than those using the original tutorial (median = 1404~s, IQR = [1293.8, 1468.8]) (Figure~\ref{fig:timeandbreak}a). 
The effect size was large (Cliff's $\delta = -0.72$, 95\% CI [$-0.96$, $-0.38$]), with a Hodges--Lehmann estimated shift of $-278.5$ seconds.

Breakdown frequency also differed significantly between conditions. 
A Mann--Whitney U test showed that participants in the AutoCue condition experienced fewer breakdowns (median = 3, IQR = [2.75, 4.25]) than those using the original tutorial (median = 7, IQR = [4.75, 8.25]; $U = 25.5$, $p = 0.007$) (Figure~\ref{fig:timeandbreak}b). 
This difference corresponded to a large effect size (Cliff's $\delta = -0.65$), with a Hodges--Lehmann estimated shift of $-3.5$ breakdowns.

To examine the potential confound of instructor assistance, we counted stuck events that required intervention.
Assistance was brief and reactive, provided only when participants could not proceed.
The Baseline condition required more interventions than the AutoCue condition (20 vs. 6 total; $Mdn=2$ vs. $0$), suggesting that AutoCue reduced the need for external recovery support.
Because instructor help accelerated recovery from stuck moments, it likely reduced prolonged delays in the Baseline condition, making the observed time benefit of AutoCue a conservative estimate.

\subsubsection{Subjective Comparison}
We used two-sided Mann--Whitney U tests for between-subject comparisons of ordinal ratings, with tie correction where applicable.
Effect sizes are reported using Cliff’s $\delta$. 
Although none of the comparisons reached the conventional significance threshold ($p < .05$), likely due to the limited sample size ($N = 12$ per condition), all three items showed consistent directional trends favoring the AutoCue-augmented condition with small-to-medium effect sizes.

\begin{figure}
    \centering
    \includegraphics[width=1\linewidth]{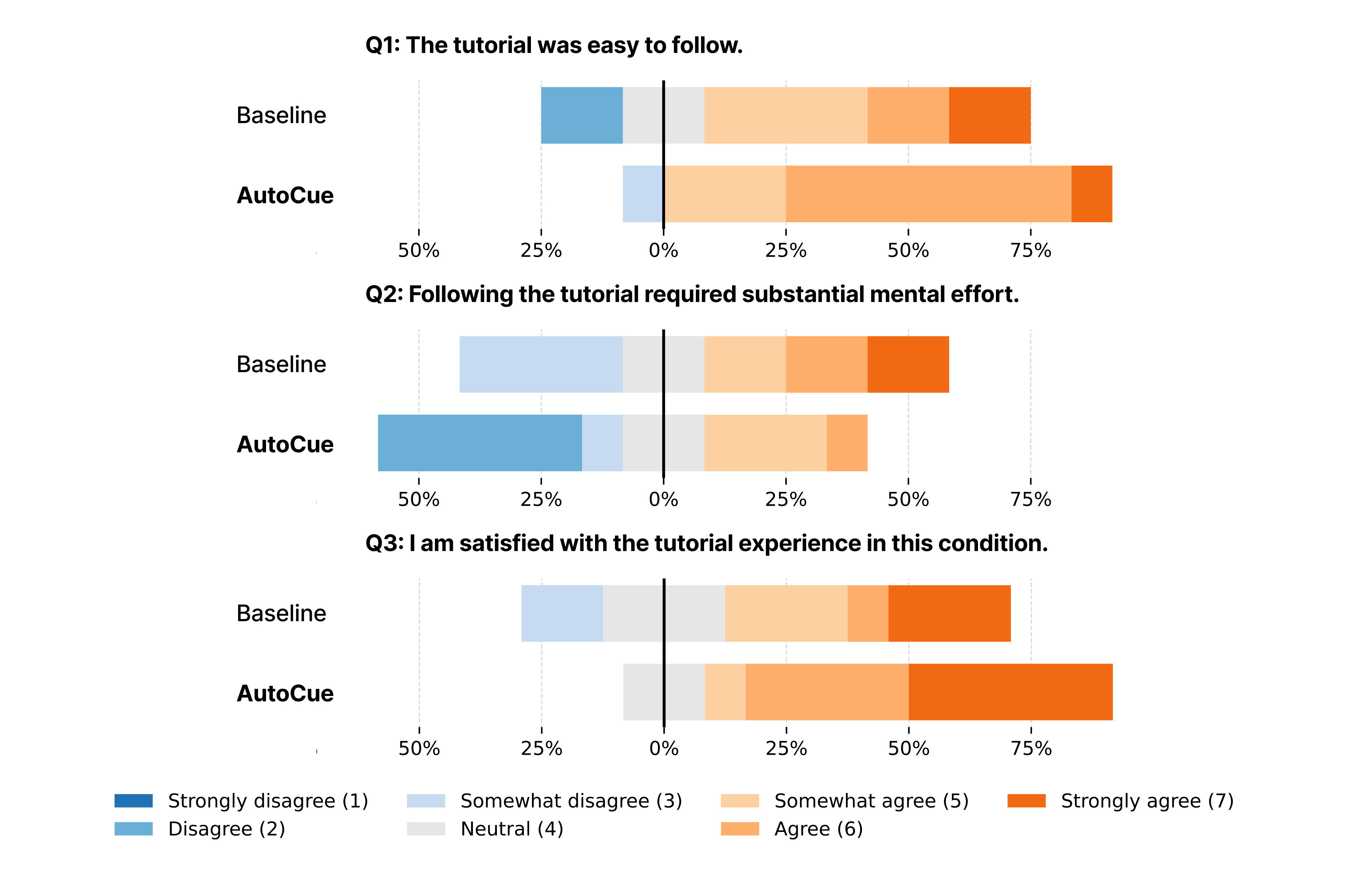}
    \caption{Distribution of 7-point Likert ratings for three tutorial experience items in the Baseline and AutoCue conditions (N=12 per condition). Bars show response percentages, and the vertical line marks the neutral point (4). For Q2, higher agreement indicates greater perceived mental effort.}
    \label{fig:Figure/Likert}
\end{figure}

\textbf{Ease of Following (Q1).}
Participants rated the AutoCue augmented tutorial as easier to follow compared to the original tutorial. 
Although the difference did not reach statistical significance (Cliff’s $\delta = 0.31, p = 0.196$), the ratings showed a consistent preference for AutoCue.

\textbf{Mental Effort (Q2).}
Participants in the AutoCue condition reported lower perceived mental effort ($Mdn = 3.5$) than those in the Baseline ($Mdn = 4.5$). 
While this difference was not statistically significant ($U = 102.0, p = 0.083$), the effect size was medium (Cliff’s $\delta = -0.42$), suggesting a practical benefit in reducing cognitive load.

\textbf{Satisfaction (Q3).}
Overall satisfaction ratings were higher for the AutoCue condition ($Mdn = 6.0$) than for the Baseline ($Mdn = 5.0$), showing a medium effect size (Cliff’s $\delta = 0.39, p = 0.102$).

\subsubsection{Does the Recall Scope Matter?}
As discussed in Section~\ref{sectionAutoCue}, AutoCue prioritizes significant interactions with clear visual feedback, using these signals to infer and visualize inputs that would otherwise remain implicit. 
Because AutoCue does not aim to cover every low-level action, we explicitly asked participants about the perceived adequacy of its recall scope. 
Most participants reported that the cue augmentation was appropriate for follow-along learning (P3, P5, P7, P9, P13, P15, P17, P19, P21). 
For example, P9 noted: “I wouldn’t change it. The detailed visual cues, like the combo for changing the view menu and the guides for dragging, are exactly what I need. 
They definitely help me understand what’s happening step-by-step.”

Overall, these responses suggest that the current recall strategy captures the interactions that matter most for learners’ execution. 
Rather than requiring exhaustive input visualization, our findings point to diminishing returns: accurate cues for key steps provide most of the benefit, whereas visualizing all inputs does not necessarily yield proportional additional gains. 
For instance, even without detecting every mouse action, highlighting the double-click required to open the \textit{Revolve} window helped participants avoid outcome mismatches that could lead to getting stuck, which happened in the baseline group (P2, P8, P10). 
Similarly, making the view-change interaction explicit (holding the spacebar while dragging the right mouse button) better aligned participants’ actions with the screencast and reduced subsequent rewinds and stuck moments observed in the baseline group (P4, P16).

\subsubsection{General Feedback on the AutoCue and Visual Design}~\label{subsectionfeedbackinvisualdesign}
When asked whether the visual cue design was helpful and to share overall feedback, most participants agreed that the visually augmented tutorial improved their learning experience by externalizing implicit input information and presenting it through our instructional cue design (P1, P3, P5, P7, P9, P13, P15, P19, P21).
Participants also provided suggestions for refining the instructional visual cue design.
P9 and P17 noted that the cues could sometimes appear too large, and that timing should be adjusted to avoid drawing attention away from the primary task ($e.g.,$ visual cue overlapped the menu content).
In addition, P1 and P17 suggested that experienced Maya users may not need the same level of support, and recommended adapting cue density or presentation to learners’ expertise rather than adopting a one-size-fits-all design. 
As P1 expressed, "I can do without the explicit cues since I already get the interaction. But to help me follow the flow, a simple visual, like a cursor trace or click rings trace, would be perfect." 

\section{Discussion and Future Work}


\subsection{Effectiveness: Externalizing Implicit Interactions}
Our formative study suggests that rewinding and getting stuck are symptoms of a deeper usability barrier. 
Many screencast tutorials leave critical input actions implicit and provide insufficient visualization for learners to interpret them efficiently.
Our user study indicates that externalizing these implicit inputs as instructional visual cues reduces interaction breakdowns and completion time in the follow-along learning process. 
For tutorials without input metadata, the AutoCue augmentation workflow surfaces and visualizes critical operations with our instructional visual cue design.
Making actions explicit reduces extraneous effort in inference and confirmation, thereby decreasing rewinds and attention switching in the learning process, which frees cognitive resources for organizing and integrating modeling logic and procedure~\cite{van2021signaling}. 
This aligns with prior research on instructional visual cues~\cite{wang2020impacts} and extends it by enabling automated cue generation for needed screencast tutorials.

\subsection{Design implications for instructional visual cue design}
Grounded in our formative findings, theoretical principles from multimedia learning and attention cueing~\cite{mayer2002multimedia,van2021signaling}, and results from our user study, we distill the following design implications for instructional visual cueing in tutorials.

\subsubsection{Action-proximal cue placement}
Consistent with our findings in Section~\ref{section:visualdesign} and prior work on reducing split attention, instructional visual cues for keyboard-modified mouse inputs (mouse + keyboard) should be presented as a bound interaction unit near the locus of action. 
Participants (P7, P11, P19, P21) reported that integrated cues helped them interpret operations and keep pace with the tutorial, especially for composite actions that combine key presses with mouse gestures. 
At the same time, co-locating cues near the cursor can occlude UI elements. 
Cue systems should therefore treat placement and transparency as configurable parameters that adapt to context.

\subsubsection{Consistency over expressiveness}
Our survey (Section~\ref{section:visualdesign}) shows that in-the-wild tutorials vary substantially in whether and how they visualize input information. 
As a result, learners must repeatedly re-learn cue conventions across creators and videos, which adds friction to follow-along learning.
Instructional visual cue design should therefore be cross-tutorial consistent over different styles and formats. 
Beyond augmenting videos without metadata, AutoCue can serve as a normalization layer that maps heterogeneous input visualizations into a consistent cue set that learners can reuse across videos, tasks, and instructors.

\subsubsection{Adaptive Scaffolding}
Our qualitative feedback suggests an expertise reversal effect. 
While most participants valued the explicit interaction indicators discussed in Section~\ref{subsectionfeedbackinvisualdesign}, an experienced Maya user (P17) occasionally described larger visual cues as "distracting."
In the interview, P17 expressed that "I don't really need all those visual cues telling me the interaction information. I'd prefer something cleaner, maybe just show me the mouse clicks position, or I can choose the format." 
Design should support adapting cue format, density, and persistence to learners' expertise and goals, instead of relying on a one-size-fits-all presentation.

\subsection{LLM-Assisted and Human-in-the-Loop Authoring}

Rather than a fully autonomous cue-generation system, AutoCue is better framed as a copilot in a human-in-the-loop authoring workflow. 
This framing aligns with HCI perspectives on AI-assisted creative and authoring tools~\cite{bourgault2025narrative}. Prior systems that leverage LLMs for tutorial and video support have focused on navigation, question answering, note generation, or tutorial restructuring~\cite{chi2025watchwithme,zhao2025noteit,yang2025videomix,chen2024tutoai}. In contrast, AutoCue focuses on a different problem: externalizing implicit interaction information that directly contributes to follow-along breakdowns in learning and repetitive manual work during tutorial authoring.

Under this framing, traceability becomes critical for accountability and effective human--AI collaboration. 
AutoCue therefore exposes intermediate evidence, including ROI regions, OCR results, and inference rationales, so creators can inspect why a cue was generated and revise it when needed.
Compared with a black-box workflow, this traceability improves transparency, supports correction, and better enables human--AI co-authoring~\cite{abdul2018trends,amershi2019guidelines}.

\subsection{Generalizability and Boundaries}


Our current implementation focuses on Autodesk Maya as a case study, but AutoCue’s modular architecture may support adaptation to other feature-rich GUI applications with observable visual feedback. 
Such adaptation would require reconfiguring application-specific operation knowledge and label mappings. 
Because AutoCue relies on constrained multimodal inference rather than application-specific trained detectors, this adaptation may be feasible without redesigning the full pipeline. 
The visual cue layer is also separable from interaction inference, which may support further customization in future deployments.

Adapting the \textit{AutoCue} pipeline to another application primarily requires modifying the Module M2's (Section~\ref{M2}) interaction menu, rather than redesigning the full pipeline.
First, the system prompt should be revised to specify the target software, allowing the multimodal LLM to perform inference with clear application-specific context.
Second, the interaction labels in the M2's interaction menu should be updated according to the target software's official documentation, including operation names, shortcut conventions, mouse actions, and keyboard-modified interactions.
This ensures that the LLM's structured output labels align with the interaction vocabulary of the target application.
Third, each interaction label should be paired with a clear description and visual cue of how the corresponding interaction is performed in the target software, grounded in official documentation (see Supplement 3.1).
These details help the LLM connect visual evidence, narration, and target-software context during inference.

Other modules, including instructional visual cue rendering, OCR extraction, and narration transcription, do not require app-specific redesign because they are designed as general components for processing screencast tutorial videos.
Therefore, we frame \textit{AutoCue} as a configurable tutorial augmentation pipeline whose software-specific layer is concentrated in the M2 interaction menu, rather than as an application-independent system that transfers without reconfiguration.

\subsection{Limitations and Future Work}
AutoCue inherits limitations from its reliance on screen-recorded visual signals. 
In the current paper, we therefore focus our strongest automatic claims on interaction events with explicit and localized visual feedback, such as UI-mediated events, which are better suited to learner-facing automatic augmentation. 
By contrast, complex state or shape changes are often expressed through more continuous and distributed visual updates, and their interpretation depends more heavily on narration, workflow context, and application-specific knowledge. 
For this reason, we do not position them as the primary target of automatic validation in this paper. 
Instead, these harder cases will be supported through the editable workflow, and a natural next step is to evaluate them more directly as an instructor-facing annotation and refinement problem, including how AI-generated candidates may reduce repetitive manual authoring effort and possible new interactive modes.

Our user study evaluation has ecological constraints. 
We used Maya as a case study in a controlled lab setting, with a single tutorial and a single task in Autodesk Maya. While this setup approximates follow-along learning, it limits conclusions about long-term retention, transfer across tasks, and generalization to diverse tutorial genres and in-the-wild learning behaviors. 
Future studies should examine longitudinal learning, multiple tutorials and tasks, and broader learner populations with varied expertise.

We also see opportunities for personalization, adapting cue density, format, and duration to learners’ expertise and preferred learning rhythm. 
Finally, a tighter human–AI co-authoring loop could allow instructors to correct cues, adjust detection preferences, and propagate edits into system behavior, enabling better alignment with author intent and goals.

\section{Conclusion}
Follow-along learning from screencast tutorials can break down when critical inputs are not captured as metadata and remain implicit in the video. 
In this work, we used formative human-participant studies to identify key sources of these breakdowns and presented \textit{AutoCue}, a multimodal LLM-assisted tutorial augmentation workflow that identifies implicit inputs in screencast tutorials and externalizes them as instructional visual cues through a structured vocabulary and grammar, producing an augmented tutorial video and an editable artifact to support human-in-the-loop completion and refinement. 
Our evaluations suggest that visualizing mouse inputs and keyboard-modified mouse inputs can reduce learners’ interaction breakdown frequency and support smoother follow-along learning, and we conclude by discussing design considerations for instructional cue systems and human-in-the-loop AI workflows, as well as opportunities to generalize this approach to other feature-rich software and tutorial domains.

\bibliographystyle{ACM-Reference-Format}
\bibliography{sample-base}
\appendix

\end{document}